\documentclass[12pt,preprint]{aastex}
\usepackage{graphicx}
\pdfoutput=1

\begin{document}

\title{Analyzing the Low State of EF Eridani with {\it Hubble Space Telescope} Ultraviolet 
Spectra\footnote{Based on observations made with the NASA/ESA Hubble Space
Telescope, obtained at the Space Telescope Science Institute, which is
operated by the Association of Universities for Research in Astronomy, Inc.,
under NASA contract NAS 5-26555, and with the Apache Point Observatory 3.5m
telescope which is owned and operated by the Astrophysical Research Consortium.}}

\author{Paula Szkody\altaffilmark{2},
Anjum Mukadam\altaffilmark{2},
Boris T. G\"ansicke\altaffilmark{3},
Ryan K. Campbell\altaffilmark{4},
Thomas E. Harrison\altaffilmark{5},
Steve B. Howell\altaffilmark{6},
Jon Holtzman\altaffilmark{5},
Frederick M. Walter\altaffilmark{7},
Arne Henden\altaffilmark{8},
William Dillon\altaffilmark{8},
Owen Boberg\altaffilmark{5},
Shannon Dealaman\altaffilmark{9},
Christian S. Perone\altaffilmark{10}}

\altaffiltext{2}{Department of Astronomy, University of Washington, Box 351580,
Seattle, WA 98195, szkody@astro.washington.edu, anjum@astro.washington.edu}
\altaffiltext{3}{Department of Physics, University of Warwick, Coventry CV4 7AL, UK}
\altaffiltext{4}{CTIO, Casilla 603, La Serena, Chile}
\altaffiltext{5}{Department of Astronomy, New Mexico State University, Box 30001, Las Cruces, N 88003}
\altaffiltext{6}{National Optical Astronomy Observatory, 950 N. Cherry Avenue, Tucson, AZ 85726}
\altaffiltext{7}{Dept. of Physics and Astronomy, Stony Brook University, Stony Brook, NY 11794}
\altaffiltext{8}{AAVSO, 49 Bay State Road, Cambridge, MA 02138}
\altaffiltext{9}{Rutgers University, Piscataway, NJ 08854; CTIO REU Student}
\altaffiltext{10}{Porto Alegre/RS, Brazil}

\begin{abstract}
Time-resolved spectra throughout the orbit of EF Eri during its low 
accretion state were obtained with
the Solar Blind Channel on the Advanced Camera for Surveys onboard the
{\it Hubble Space Telescope}. The overall spectral distribution exhibits peaks at 1500
and 1700\AA, while the UV light curves display a quasi-sinusoidal modulation over the
binary orbit. Models of white dwarfs with a hot spot and cyclotron emission
were attempted to fit 
the spectral variations throughout the orbit.
A non-magnetic white dwarf with
a temperature of $\sim$10,000K and a hot spot with central temperature of 15,000K
generally matches the
broad absorptions at
1400 and 1600\AA\ with those expected for the quasimolecular H features
H$_{2}$ and H$_{2}^{+}$. However, the flux in the core of the Ly$\alpha$ absorption
 does not go to zero, implying
an additional component, and the flux variations throughout the orbit are not
well matched at long wavelengths.
Alternatively, a 9500K white dwarf 
with
a 100 MG cyclotron component can fit the lowest (phase 0.0) fluxes, but the 
highest fluxes (phase 0.5) 
require an additional source of magnetic field or temperature. The 100 MG field
required for the UV fit is much higher than that which fits the optical/IR 
wavelengths, which would support previous suggestions of a complex field structure in polars.
\end{abstract}

\keywords{binaries: close --- novae, cataclysmic variables --- stars:
individual (EF Eri) ultraviolet: stars}

\section{Introduction}

EF Eri is a well-studied cataclysmic variable that contains a highly magnetic
white dwarf. It was identified in 1979 as the fourth known
Polar (Griffiths et al. 1979; Hiltner et al. 1979; Tapia 1979) with an
81 min orbital period. During the next 20 years, it remained in a
bright high state near 14th mag, with highly modulated optical, IR and X-ray light
curves due to a high mass transfer rate and X-ray and cyclotron emission from the
accretion region (Bailey et al. 1982). Models of the cyclotron humps determined
magnetic field strengths of 16.5 and 21 MG (Ferrario, Bailey \& Wickramasinghe
1996). However, in 1997, EF Eri entered a low state near 18th mag 
(Wheatley \& Ramsay 1998)
which lasted for 9 years, with only short excursions to a high state since
2006. Low states are advantageous to study the underlying white dwarf and
secondary star without the effects of the accretion flow. In the low state,
the optical spectrum of EF Eri shows broad Balmer absorption lines with
Zeeman splitting indicating a 14 MG field (Wheatley \& Ramsay 1998). 
Beuermann et al. (2000) modeled the optical spectrum
with a 9500K white dwarf and a 15,000K hot spot over 6\% of its surface.
As there were no features from a secondary star, they concluded that the
spectral type of the companion must be later than M9. 
{\it Spitzer} mid-IR emission suggested the
companion was an L or T dwarf (Howell et al. 2006b). Howell et al. (2006a) used the 
H$\alpha$
emission from this substellar secondary to produce a radial velocity curve,
which provided a spectroscopic ephemeris as well as a mass estimate for the
secondary of 0.055 M$_{\odot}$ for a white dwarf mass of 0.6 M$_{\odot}$.
Thorstensen (2003) attempted parallax measurements but the faintness of the
source and the assumptions used
yielded
a small range of distances centered near 113 pc or a larger range near 163 pc.

While the accretion rate in low states is reduced by several orders of
magnitude from the high state, and there is no evidence of a stream of
mass transfer during the low state of EF Eri, UV studies have shown that
hot spots with temperatures of 30,000-70,000K are likely present on
the white dwarfs of Polars during low states.  
{\it IUE} observations of AM\,Her in the low state revealed an
orbital-phase dependent flux modulation, which
Heise \& Verbunt (1988) interpreted as the signature of moderate
heating of a large fraction of the white dwarf
surface. G\"ansicke et al. (1995) developed this hypothesis further,
showing that a large pole cap covering $\sim10\%$ of the white
dwarf surface is present both during the high state and the low state,
and that, at least in the case of AM\,Her, the luminosity of this
heated pole cap is comparable to that of the hard X-ray and cyclotron
emission from the post-shock region, solving the so-called ``soft
X-ray puzzle'', i.e. the soft X-ray excess with respect to the simple
reprocessing model of Lamb \& Masters (1979) and
King \& Lasota (1979). Evidence for large, ``warm'' pole-caps was
found in many other polars (e.g. Stockman et al. 1994;
 G\"ansicke et al. 2000, Schwope et al. 2002, Araujo-Betancor et al 2005).
G\"ansicke et al (1998) developed a 3-dimensional model of a white
dwarf with a heated pole cap to analyze high-time resolution
\textit{HST}/GHRS observations of AM\,Her in a high state, and
\textit{FUSE} and \textit{HST}/STIS observations of AM\,Her in a low
state (G\"ansicke et al 2006), leading to more precise constraints
on the size, temperature, and location of the heated pole
cap. Especially for the low-state data, where no emission from the
accretion stream was present, this model provided an excellent fit to
both the observed modulation of the ultraviolet flux, as well as to
the changes in the Lyman line profiles. 

{\it GALEX} photometric observations of EF Eri during its low state 
(Szkody et al. 2006; 2008) in two broad band filters centered near 1550\AA\
and 2300\AA\ were modeled with a 9500K white dwarf and a spot of radius 5.5$^\circ$ 
with a temperature
of 24,000K. Schwope et al. (2007) analyzed archival
{\it XMM-Newton} observations to detect an X-ray flux of 6$\times$10$^{-14}$
erg cm$^{-2}$ s$^{-1}$ during the low state, corresponding to an x-ray
luminosity of 2$\times$10$^{29}$ erg s$^{-1}$. While the specific accretion
rate would then be only 0.01 g cm$^{-2}$ s$^{-1}$, implying no
accretion shock, they suggested the X-rays come
from some residual accretion at the poles. Li et al. (1994, 1995) have
produced models for high field white dwarfs where the white dwarf can capture
all the stellar wind from the secondary and funnel it to the poles.

Further support for some accretion during the low state
comes from the presence of cyclotron humps
in the IR (Harrison et al. 2004; Campbell et al. 2008). 
Campbell et al. (2008) 
explored whether
high magnetic field cyclotron emission could be producing the {\it GALEX} 
UV variations as well as the IR. 
The high field (240 MG) system
AR UMa does show UV variability that is likely due to cyclotron emission (G\"ansicke et
al. 2001) and QS Tel showed humps at 2200\AA\ and 2430\AA\ that could be interpreted
as cyclotron humps in a 65 MG field (Rosen et al. 2001).
The models of Campbell et al. (2008)
showed that the UV variations in EF Eri could be cyclotron if the field
was 115 MG, while the IR variations require a field of 13 MG. Some
confirmation of the high field required for the UV is evident from the
work of  
Beuermann et al. (2007), who used
Zeeman tomography and determined that EF Eri has
a complex field with some regions as high as 100 MG. 

In order to better understand the cause of the UV light and variability,
we obtained time-resolved spectroscopy with the {\it Hubble Space Telescope (HST)}.
 These data and our modeling efforts with a heated polecap and cyclotron emission
are described below.
 
\section{Observations}

The Solar Blind Channel (SBC) on the Advanced Camera for Surveys (ACS)
was used with prism PR110L throughout four sequential 
{\it HST} orbits on 2008 January 17. This prism provides wavelength
coverage with useful sensitivity from $\sim$1200-1900\AA, the prism
results in non-linear resolution from $\sim$2\AA\ pixel$^{-1}$ at the blue
end to $\sim$40\AA\ pixel$^{-1}$ at the red end. The ACCUM mode was used
with continuous exposures of 239 sec during each satellite orbit. The
first orbit had 9 exposures (due to the setup time) while the remaining
3 orbits contained 10 exposures. The setup exposure on the target was
accomplised with the F140LP filter and an integration time of 15 s.
The observation times for each orbit are summarized in Table 1.

The reduction package aXe1.6 (Kuemmel et al. 2009) provided by STScI was
used to extract the target and produce a flux and wavlength calibrated
spectrum for each 239 sec exposure. To 
obtain the best flux level, we used an extraction width of $\pm$17 pixels
corresponding to $\pm$ 0.5 arcsec. To show the variability, we
created light curves
in six bands (1220-1260\AA, 1330-1380\AA, 1430-1530\AA, 1530-1640\AA, 
1640-1720\AA, and 1780-1820\AA) by integrating the fluxes within these regions.
Phases were calculated using the spectroscopic phasing of Howell et al. (2006a).

To insure that EF Eri was in its low state during the {\it HST} observations,
we solicited optical ground data from AAVSO members and observatories
in the Northern and Southern hemisphere in the nights preceding and
following the HST times. The New Mexico State University (NMSU) 1m telescope
provided three nights of measuremnts with UBVR filters and time-resolved
differential light curves in V filter. The Small and Moderate Aperture Research
Telescope System (SMARTS) obtained B band photometry with a CCD on the
0.9m telescope. Spectra throughout an orbit were obtained on the 3.5m
telescope at Apache Point Observatory (APO) using the Dual Imaging
Spectrograph (DIS) with the high resolution gratings (resolution $\sim$2\AA)
that provided simultaneous spectra from 3900-5100\AA\ and from 6300\AA\ to
7300\AA. The times of the optical data close to the HST observation are listed
in Table 1.

\section{Optical Results}

Both the optical photometry and spectroscopy showed that EF Eri remained
in its deep and extended low state during the {\it HST} observations. The
calibrated data provided B=18.29, V=18.40 and R=18.10 the night preceeding
the {\it HST} observation. The
SMARTS photometric point from that night is shown in Figure 1 as a star superposed on
the B light curve accumulated on 68 nights from 2007 June 20 to 2008 March 23 while
EF Eri was in a low state (Walter 2009). The NMSU V light curves
from the nights preceding and following the {\it HST} times
 are shown in Figure 2. Both
light curves show the normal low state optical modulation with a steeper rise
to a peak brightness near phase 0.5 and a slower decline to minimum near
phase 0.0. While the data from January 18th show EF Eri slightly fainter, the larger
error bars on this night are within the range of variation evident in Figure 1.
The APO spectra (Figure 3) show the typical low state features of broad Balmer absorption
lines from the white dwarf flanking weak emission (Wheatley \& Ramsay 1998).
Howell et al. (2006a) determined that the source of the Balmer emission is the
substellar secondary and is due to stellar activity. The level of emission apparent
in Figure 3 is somewhat less than that shown in Fig 2 of Howell et al. (2006b)
during 2006 January but within the range of spectra shown in that paper during 2005.

\section{{\it HST} Spectra and Light Curves}

The individual {\it HST} orbits covered about 0.4-0.5 of the orbital phase of
EF Eri on each pass, with all phases covered at least once in the 4 orbit
sequence. Figure 4 shows the variability of the flux
within each of the four {\it HST} orbits. In this plot, each 239s integrated 
flux point was
divided by the average flux, and then converted to a magnitude scale.
Due to the large variability throughout the binary orbit of EF Eri,
we averaged the spectra 
in three ways, which are shown in Figure 5. The solid line is the
average of all the data, the top dashed line is the average of the 9
integrations within 23\% of the peak of the light curves (the highest
 circled points
shown in Figure 4), and the bottom dotted line is the average of the 9 points
within 23\% of the minimum flux of the light curves (the lowest circled points 
in Figure 4). The points were selected to avoid the rising and falling portions of the 
light curve. All three averages show the lowest flux at Ly$\alpha$, a
broad minimum near 1600\AA, and peaks near 1500 and 1700\AA.  While
these features are preserved in all the averages, the heights of the
peaks and the depths of the 1600\AA\ absorption feature change between peak and
trough spectra. The 1500\AA\ peak becomes less prominent compared to 
the 1700\AA\ peak (which causes a corresponding decrease in the depth
of the 1600\AA\ absorption feature) during the minimum flux times.

It is possible to interpret these spectra in two ways. If the
underlying flux is primarily from a cool white dwarf, the 1600\AA\
absorption can be identified as quasi-molecular hydrogen H$_{2}$ which
is present in white dwarfs with temperature $\leq$13,500K (Koester et al.
1985). There is also some evidence for quasi-molecular H$_{2}^{+}$ blueward pf
1400\AA, which is present in white dwarfs cooler than 20,000K.
Alternatively, if the UV flux comes primarily from cyclotron, the
humps at 1500 and 1700\AA\ could be caused by cyclotron harmonics.
Support for this origin is the steep decline longward of $\sim$1750\AA,
which is different from the flatter continuum distribution of a cool white
dwarf.

To determine if the drop in
flux at longer wavelengths could be a
calibration problem, we extracted the SBC spectra of two white dwarf
standard stars (WD1657+343 and LTT9491) taken with the PR110L and PR130L
prisms from the {\it HST} archive. The hot white dwarf WD1657+343 showed a
declining spectrum in both prisms from 1250-1850\AA\ while
the DZ white dwarf LTT9491 showed a discrepancy longward of 1800\AA,
with the PR110L spectrum being lower than the 
PR130L spectrum by 5\% at 1850\AA. To further check, we extracted two {\it IUE} spectra
of LTT9491 from the {\it IUE} archive (shown in Figure 6 along with the SBC spectrum).
This also shows that the SBC calibration is good at short wavelengths but has
problems longward of 1750\AA.
Thus, we regard the calibrated fluxes of EF Eri to be accurate only up to 1750\AA.

For further phase resolution with good S/N, we
created light curves as a function of orbital phase by integrating the
spectra over the 6 UV bandpasses described in Section 2.
These light curves are shown in Figure 7. The bandpasses were chosen
to span different features of the spectra. The 1220-1260\AA\ bandpass, which
covers the core of Ly$\alpha$, and the 1780-1820\AA\ one, which is the long
wavelength downturn in flux,  show the least change as a function of
the orbit. The bandpasses that include the 1500\AA\ peak (1430-1530\AA)
and the 1600\AA\ absorption feature (1530-1640\AA) have the largest amplitude
variablity (over a factor of 2) with peak flux occuring near the same
phases as the optical variability (Figures 1 and 2). However, in
contrast to the optical, the UV amplitudes are larger and do not show
the same asymmetry in shape that is evident in the optical light.

We attempted to model the spectra and the light curves with hot spots
on a white dwarf as well as with cyclotron components.
 
\section{Spot Model}

Using a similar approach to our modeling efforts for AM Her (G\"ansicke et al.
1998, 2006), we used the same code to fit 
the ACS/SBC light curves of EF\,Eri. In
brief, the surface of the white dwarf is represented by small tiles,
with the temperature of each being adjustable. The emission of each
tile is described by a white dwarf model spectrum with the
corresponding temperature, and the total emission of the white dwarf
is computed by integrating the contribution of all tiles on the visible
hemisphere. To keep the number of free parameters small,
  the heated region is
represented by a circular spot with an opening angle $\theta_{\rm spot}$, located at
a colatitude $\beta_{\rm spot}$ with respect to the rotation axis of the white
dwarf, and an $\psi_{\rm spot}$ with respect to the axis connecting the white
dwarf and the donor star. The temperature distribution within the spot
is assumed to drop linearly in angle from $T_{\mathrm{cent}}$ at the center of the
spot to the temperature of the unheated white dwarf, $T_{\mathrm{wd}}$, at the edge
of the spot. Koenig et al. (2006) modeled cyclotron heating in the case of
AM Her, and found a spot with a flat temperature near
 the innermost 10$^{\circ}$, 
which dropped off approximately linearly out to $\sim$35$^{\circ}$. For
modeling the light curves of EF Eri, a linear drop is the simplest assumption
with the smallest number of parameters. 
Additional parameters are the white dwarf radius $R_{\mathrm{wd}}$,
and the distance to the system $d$. We simultaneously fitted the
ACS/SBC light curves in wavelength ranges 1330--1380\,\AA,
1430--1530\,\AA, 1530--1640\,\AA, and 1640--1720\,\AA. The
1220-1260\,\AA\ and 1720--1820\,\AA\ light curves were omitted from
the fit as they were subject to excessive noise and problems in the
flux calibration, respectively. Free parameters in the fit were $T_{\mathrm{eff}}$,
$R_{\mathrm{wd}}$, $T_{\mathrm{cent}}$, $\theta_{\rm spot}$, $\beta_{\rm spot}$, and
 $\psi_{\rm spot}$; we kept the distance fixed
at $d=120$\,pc and the inclination of the binary orbit at
$i=50^{\circ}$. The free parameters were adjusted using an evolution
strategy (Rechenberg 1994). It carries out a sequence of mutations of the 
parameters and
evaluates them against a fitness function, which we chose to be simply
$\chi^2$. The step size of the mutation is itself optimized as a
function of the convergence of the evolution process.
Starting the fit with different
initial parameters led to consistent convergence in the
six-dimensional parameter space. We list in
Table 2 the average values and standard
deviations of six solutions found from very different initial start
parameters. 

Inspecting the best-fit light curves to the ACS/SBC observations
(Figure 8) reveals that the
1330--1380\,\AA\ data are reasonably well fitted, but that the model
significantly over/under predicts the observed fluxes in the three
longer wavelength bands. This contrasts with the quality of the fit
achieved for the low-state observations of AM\,Her
(G\"ansicke et al. 2006). A possible explanation is the presence of
some residual ultraviolet flux from low-level accretion, as suggested
by the fact that the flux in the core of Ly$\alpha$ (Figure 5) does
not drop to zero, as expected for the low temperatures of the white
dwarf and its pole cap. 
However, while the model fails to exactly match the observed fluxes in
the four wavelength ranges, it reproduces well the overall spectral
shape of the ultraviolet spectra of EF\,Eri as a function of the
orbital phase, i.e. as a function of the geometric projection of the
heated pole cap (Figure 8). In particular, the
broad depression seen at all phases near 1600\,\AA\ corresponds to the
quasi-molecular H$_{2}$ absorption that occurs in cool, high-gravity
atmospheres (Koester et al. 1985, Nelan \& Wegner 1985), and the steep
rise in flux near 1400\,\AA\ observed at orbital maximum is consistent
with the quasi-molecular H$_{2}^{+}$ absorption depressing the flux shortward of
1400\AA. There is some uncertainty
as to how the magnetic field affects the lines and quasimolecular features in
the spectrum of a white dwarf. G\"ansicke et al. (2001) have shown that the
UV flux distribution of the highest magnetic field polar AR UMa differs greatly
from that of non-magnetic white dwarfs in having a flatter UV flux distribution
compared to longer optical wavelengths. Overall, the low-state
ultraviolet spectrum of EF\,Eri is rather similar to that of VV\,Pup
($T_{\mathrm{eff}}=12300$\,K, Araujo-Betancor et al. 2005), which also
exhibits the 1400\,\AA\ and 1600\,\AA\ absorption features. The
1600\,\AA\ absorption feature disappears in somewhat hotter atmospheres, whereas
the 1400\,\AA\ feature remains visible up to $\simeq20\,000$\,K
(Araujo-Betancor et al. 2005, G\"ansicke et al. 2006). 

The geometry of
the heated pole cap is illustrated in Figure 10.
The white dwarf temperature of $\simeq10000$\,K is consistent with the
results from the analyses of optical low-state spectroscopy
(Beuermann et al. 2000). The white dwarf radius corresponds to a
mass of $\simeq0.8M_\odot$, but depends on the assumed distance~--~a
true distance larger (lower) than 120\,pc would imply a
correspondingly larger, lighter (smaller, heavier) white dwarf. The
maximum temperature of the spot, $T_{\mathrm{cent}}\simeq15000$\,K is somewhat
lower than the value estimated by Schwope et al. (2007), which was
based on the analysis of the orbital-averaged spectral energy
distribution of EF\,Eri as observed with \textit{XMM-Newton}. Azimuth
and co-latitude of the heated pole cap are broadly consistent with the
location of the main accretion region in EF\,Eri as determined by
Beuermann et al. (2007) from low-state Zeeman tomography. The
fractional area of the heated pole cap in EF\,Eri is $\sim22$\%, which
is larger than that in AM\,Her (G\"ansicke et al. 2006), but
comparable to those in V834\,Cen, BL\,Hyi, MR\,Ser, and V895\,Cen
(Araujo-Betancor et al. 2005).

\section{Cyclotron Model}

The cyclotron models were produced using a Constant Lambda (CL) code, which is 
given a more complete overview in Campbell et al (2008). The underlying 
algorithm is based on the assumption that the plasma in the accretion column 
exists in the large Faraday rotation limit $\phi >> 1$, where 
$\phi=\frac{e^{3}\lambda^{2}}{2\pi m^{2}c^{4}} \int_{0}^{s}n_{e} \vec B(s) ds$.
 Radiative transfer therefore decouples into two distinct magnetoionic modes: 
the ordinary (o) and extraordinary(e), whose optical depths are given by 
$\tau_{o,e} = \Lambda \phi_{o,e}$, where $\Lambda= l\omega^{2}_{p}/(c\omega_{c})$, 
where $l$ is the path length through the plasma, $\omega_{p}$ is the plasma 
frequency, $\omega_{p}= (4\pi Ne^{2}/m)^{1/2}$, and $\omega_{c}$ is the 
frequency of the cyclotron fundamental, $\omega_{c}=eB/mc $, with $m$ being 
the relativistic mass of the gyrating particles. 

Thus, through the parameter $\Lambda$, it is possible to scale a dimensionless 
emitting slab to approximate the observed characteristics of a true 3-D 
cyclotron emission region. However, the model also requires three additional 
input parameters to calculate the absorption coefficients $\phi_{o,e}$: B, 
the magnetic field strength, $\Theta$, the angle from the magnetic field line 
to the line-of-sight of the observer, and kT, the global isothermal temperature
 of the emitting slab. 

Due to the large size of the 4-D parameter space, and the computational effort needed
to enumerate and evaluate all points of the search space, an automated 
optimization routine is required. We used a Genetic Algorithm (GA) to identify
the optimal solution (for a good primer
 on genetic algorithms, see Charbonneau (1995)).  
GAs are a class of optimization algorithms inspired by biological evolution, these
algorithms encode potential solutions on a chromosome-like data structure (called
 ``individuals'') and apply 
simulated genetic operators to these structures in order to preserve the essential
 information. GAs begin by establishing a
random population of individuals, in our case, always respecting the interval of 
the 4-D search space. These individuals then
evolve through generations by processes of recombination (crossover) and mutation,
 the natural selection acts at
each generation by usually prioritizing individuals with higher scores (most well
-adapted), given by the fitness function.
While several publicly available GAs frameworks exist, we chose to use Pyevolve
 for our implementation. Pyevolve is an
open-source, modular and object-oriented framework for Evolutionary Computation 
written in Python language by Christian S. Perone (Perone 2009).
Pyevolve is both modular and highly extensible, containing several independent 
selection schemes which alter how the fitness function is used to populate the 
next generation. 

In our work, the individual was constructed out of the four CL parameters with 
variants of each parameter randomly selected from within the region defined by 
[ 65.00 $\leq$B(MG) $\leq$ 125.00, 0.1 $\leq$kT(keV) $\leq$ 25,  
0.1 $\leq$log$\Lambda$ $\leq$ 8, 25$^{\circ}$ $\leq$$\Theta $$\leq$ 85$^{\circ}$]. 
A cyclotron model was then run for every individual in the population, and then
 co-added with a white dwarf (WD) at 120 pc and normalized so that the area 
under the composite model equals the area under an observed spectrum at a 
specified phase($\Phi$ = 0.00). To enable direct comparison with the hotspot 
results, the WD was computed by using the G\"ansicke formulation discussed in 
the previous section, but at a lower temperature of 9500 K. The uncertainty
in the temperature is about 1000K around a value of 9750K (Schwope et al. 
2007). Since the grid of white dwarf models in our cyclotron fitting had
a spacing of 500K, the 9500K point was chosen as closest to the Schwope
et al. (2007) value. Our fitness 
function ranked each individual by looking for the minimum the value of 
$\chi^{2}_{\nu}$, normalized to the entire population.  We employed a Roulette 
wheel selection scheme, using the Uniform Crossover method with a rate of 90\%
and a Real Gaussian Mutation method with a rate of 15\%.
 It should be noted that convergence to global 
minimum requires maintaining genetic diversity over many generations. To this 
end, a large population of 250 individuals were run against the  $\Phi$ = 0.00 data 
spectrum over 250 generations, finding an optimal solution of 
$\chi^{2}_{\nu}$ = 1.24 of B = 100 MG, kT = 6.54 keV, log$\Lambda$ = 6.88, 
and $\Theta$ = 51.26, which took nearly two weeks of computational resources to
evaluate.

The 39 individual observed spectra of EF Eri 
 were co-added within 0.10 phase bins (with about four spectra within each bin)
 for comparison to the cyclotron models. Orbital photometry was estimated by 
integrating these data through six predefined bandpasses. Using an 
independently determined orbital inclination of $i$ = 58$^{\circ}$, 
(Campbell et al.  2008), and the value of $\Theta$ at cyclotron minimum, 
$\Theta$ = 51.26$^{\circ}$, is sufficient to constrain the magnetic 
co-latitude to $\beta$ = 6.7$^{\circ} $.  At every other phase-point, a new 
cyclotron model was computed keeping B, kT, and log$\Lambda$ constant while 
allowing $\Theta$ to vary in accordance with the orbital geometry determined above. 
We then integrated the resulting spectra through each band to produce our 
lightcurves. Figure 11 shows the optimized parameter set that was determined and
 the derived quantities while Figure 12 shows cyclotron emission as a function 
of the viewing angle (the cyclotron beam), with the total flux representing the
 area under the curve shown. The best-fit model at $\Phi$ = 0.00 is shown in 
Figure 13 (top). This fit can reproduce the minimum spectra well with B=100 MG,
 kT=6.54 keV, with $\chi^{2}_{\nu}$ = 1.24. However, the corresponding 
solution at cyclotron 
maximum produces a rather poor fit to the
 observed spectrum at that phase (Figure 13; bottom). The fit to all 10 phases 
is shown in Figure 14 as a stacked series of spectra, and the integrated light 
is similarly represented as the light curves for 6 bandpasses in Figure 15. 
From all possible fits within a large parameter space, we have ruled out a 
single magnetic component, fixed spot model that can explain the light during 
the entire orbit. Either variable magnetic field strengths or more than one 
cyclotron spot would be needed to account for the variability observed.
Given that our use of constant $\Lambda$ and kT applied to the whole
shock is a simplistic representation of a shock that could have
 a variety of temperatures,
densities and/or magnetic fields, our work is just a first attempt at the problem.  

To further ensure that the Pyevolve was converging on the optimal solution, we also ran
 a large batch of models for B = 100 MG in an evenly spaced grid stepping by
 ($\Delta$kT, $\Delta$log$\Lambda$) = (1.0,1.0) over the ranges of kT= [0:20] 
and log$\Lambda$ = [0:8], respectively. The  $\chi^{2}_{\nu}$ value for each 
were then computed at cyclotron minimum, cyclotron maximum, and also averaged 
over 10 phase-points throughout the orbit. The data were interpolated and 
mapped to a color scheme to produce heat maps shown in Fig. 16. To orbitally 
modulate the light-curve, we again varied $\Theta$ to be consistent with an 
inclination of $i$ = 58$^{\circ}$, but computed each set of 
heat maps using various magnetic co-latitudes 
($\beta$ = 1, 2, 3 $\dots$ 20$^{\circ}$) as inputs.  $\beta$ = 7$^{\circ}$ was 
found to be the optimal solution, where the orbitally averaged heat map had the
lowest $\chi^{2}$. A quick inspection of Fig. 16 demonstrates 
that the Pyevolve is working. Not only has the program converged to the best 
model at cyclotron minimum (the white spot near kT = 6.54, log$\Lambda$ = 6.88),
but an identical solution produces the best model at cyclotron maximum, and 
indeed averaged through the orbit. Interestingly, while cyclotron minimum has 
several well defined $\chi^{2}_{\nu}$ minima, the orbital average has no clear 
best model, presumably because no model fits particularly well.

Finally, we computed the cyclotron flux, F$_{cyc}$, luminosity, L$_{cyc}$, and 
specific mass accretion rate, $\dot{m}$.  The following recipe was undertaken: 
(1) Calculate the accretion spot size. Essentially, this is the ratio of the 
observed cyclotron flux and the raw model intensity both integrated over all 
wavelengths and subsequently diluted by the surface area of a sphere of radius 
120 pc.  Normalization to the data was accomplished by scaling to 
I$_{\lambda}({\rm Data})$ which is the wavelength integrated model intensity 
normalized so that the integrated light over the range 
(1220 $\leq \lambda_{{\rm Data}}\leq$ 1820) equals the integral of the observed
 fluxes. I$_{\lambda}({\rm Model})$ is just the integrated raw model intensity.
 (2) The integrated cyclotron flux was calculated by determining the model 
intensities at all polar angles, and integrating through both angle and 
wavelength. (3) L$_{Cyc}$ is the product of the cyclotron flux and the spot 
size. (4) The specific mass accretion rate was computed assuming a 
0.7 M$_{\odot}$ WD, with R$_{WD}$ determined using the Nauenberg (1972) WD mass-radius
 relationship. The process is also described numerically below.

\begin{eqnarray}
		 A_{spot}&=&\frac{\int_{0}^{\infty} I_{\lambda}({\rm Data}) {\rm d}\lambda}{ \int_{0}^{\infty} I_{\lambda}(\rm{Model}) {\rm d}\lambda}\Big(1.20 \times 10^{38} d^{2}_{120} \rm{cm}^{2}\Big) \\
	          F_{cyc} &=& 2\pi\int_{0}^{\frac{\pi}{2}}\int_{0}^{\infty} I_{\lambda}(\Theta) \sin\Theta {\rm d}\theta {\rm d}\lambda\\
	          L_{Cyc}& = &F_{cyc}A_{Spot} \\
		\dot{m} &\approx& \frac{L_{cyc}R_{WD}} {GM_{WD}}
\end{eqnarray}

From our final results (Figures 13-15), it is obvious that a single cyclotron component 
can only be fitted to the HST 
data for a small range in phase (near $\phi$ = 0.00). To fully explain the 
dataset, it is likely that either a second component is necessary, or the 
addition of a hotspot is needed. We are reluctant to explore the implications 
of a second cyclotron component without the inclusion of low-state polarimetry 
to assess the degree to which the UV emission of EF Eri is polarized.

\section{Conclusions}

Our low resolution {\it HST} spectra have revealed the first UV spectrum of
EF Eri during its low accretion state and the corresponding changes in
the spectra throughout its orbit. Two interpretations
of the lowest orbital UV fluxes at phase 0.0 are possible. If most
of the flux is from a white dwarf at a distance of 120 pc, the UV spectrum can be matched with
a temperature of $\sim$10,000K and a hot spot with central temperature of 15,000K (similar values found by Beuermann et al. (2000) for the optical region).
The broad absorptions at
1400 and 1600\AA\ match those expected for the quasimolecular H features
H$_{2}$ and H$_{2}^{+}$ but the flux in the core of Ly$\alpha$ does not go to zero, implying
an additional component. On the other hand, the UV spectrum can also
be adequately modeled with a 9500K white dwarf at 120 pc that contributes 
longward of 1600\AA\
in combination with a high (100MG) cyclotron component dominating at the 
shortest
wavelengths. 

Both models have difficulty reproducing the observed orbital flux variations,
especially for wavelengths longer than 1400\AA.
A normal white dwarf with a hot spot does not produce enough amplitude near
1500\AA\, but produces too much flux at 1700\AA\ to account for the phase 0.5
modulation. This type of model has problems producing the amplitudes of the 
optical flux variations as well. 
The cyclotron models also require some additional component (either in temperature
or in additional field strengths) to account for the
phase 0.5 UV fluxes longward of 1400\AA. The failure of the simple Constant
Lambda model to fit all phases points to the need for more complex
multi-temperature, multi-lambda models. 
While our model excludes a simple UV
emitting pole, a more complex accretion region is very possible. The Zeeman
tomography work by Beuermann et al (2007) on EF Eri, BL Hyi and CP Tuc 
shows that the field structures are not
simple dipoles and there is a range  of field strengths in all three of
these polar systems.
For EF Eri, the multipole fields range from  4 to 111 MG.  This range could
accomodate the fields of 13-21 MG that optimally fit the optical and IR spectral
regions (Ferrario et al. 1996; Campbell et al. 2008) as well as the UV field
strength estimated from ${\it GALEX}$ (Szkody et al. 2008) and our HST data. 
An additional uncertainty
is related to the distance. 
Ultimately, the choice of
model would be best served by UV spectropolarimetry to define the cyclotron
component correctly, but a good parallax and UV spectra can pin down the
UV luminosity from a high field accreting white dwarf. The theoretical
profiles for Ly$\alpha$ and the quasimolecular H features at high fields
are not well-known, so observations are currently the best way to determine
the influence of the field on the resulting radiation. 

\acknowledgments

We gratefully acknowledge the help of the AAVSO members who provided magnitude
estimates close to the time of the {\it HST} observations. Support for this
research was provided by STScI grant HST-GO-11162.01-A.

\clearpage

\begin{deluxetable}{lllcc}
\tablewidth{0pt}
\tablecaption{Observation Summary}
\tablehead{
\colhead{UT Date} & \colhead{Observatory} & \colhead{Instrument} & \colhead{Time} & \colhead{Int (s)} }
\startdata
2008 Jan 15 & NMSU & CCD V filter & 02:32-05:34 & 300 \\
2008 Jan 16 & NMSU & CCD V filter & 01:27-03:09 & 300 \\
2008 Jan 16 & APO & DIS & 01:48-03:30 & 600 \\
2008 Jan 16 & SMARTS & ANDICAM B & 02:02:28 & 100 \\
2008 Jan 17 & {\it HST} & SBC & 10:25-11:01 & 239 \\
2998 Jan 17 & {\it HST} & SBC & 11:59-12:39  & 239\\
2008 Jan 17 & {\it HST} & SBC & 13:35-14:14  & 239 \\
2008 Jan 17 & {\it HST} & SBC & 15:11-15:50  & 239 \\ 
2008 Jan 18 & NMSU & CCD V filter & 01:31-04:22 & 300 \\ 
\enddata
\end{deluxetable}

\clearpage
\begin{deluxetable}{lc}
\tablecolumns{2}  
\tablewidth{0pc}  
\tablecaption{Best-fit Parameters for 
 White Dwarf plus Hot Spot Model}
\tablehead{  
\colhead{Parameter} &
\colhead{Value} }
\startdata  
$T_{\mathrm{eff}}$   & $9850\pm175$\,K \\
$T_{\mathrm{cent}}$  & $15095\,\pm66$\,K \\
$R_{\mathrm{wd}}$    & $(7.95\pm8)\times10^8$\,cm\\
colatitude ($\beta_{\rm spot}$)  & $17.4\pm1.4^{\circ}$ \\
opening angle ($\theta_{\rm spot}$) & $57.2\pm2.2^{\circ}$ \\
azimuth   ($\psi_{\rm spot}$) & $4.1\pm0.4^{\circ}$ \\
\enddata  
\end{deluxetable}  

\clearpage
\begin{figure} [h]
\figurenum {1}
\includegraphics[angle=-270,width=16cm]{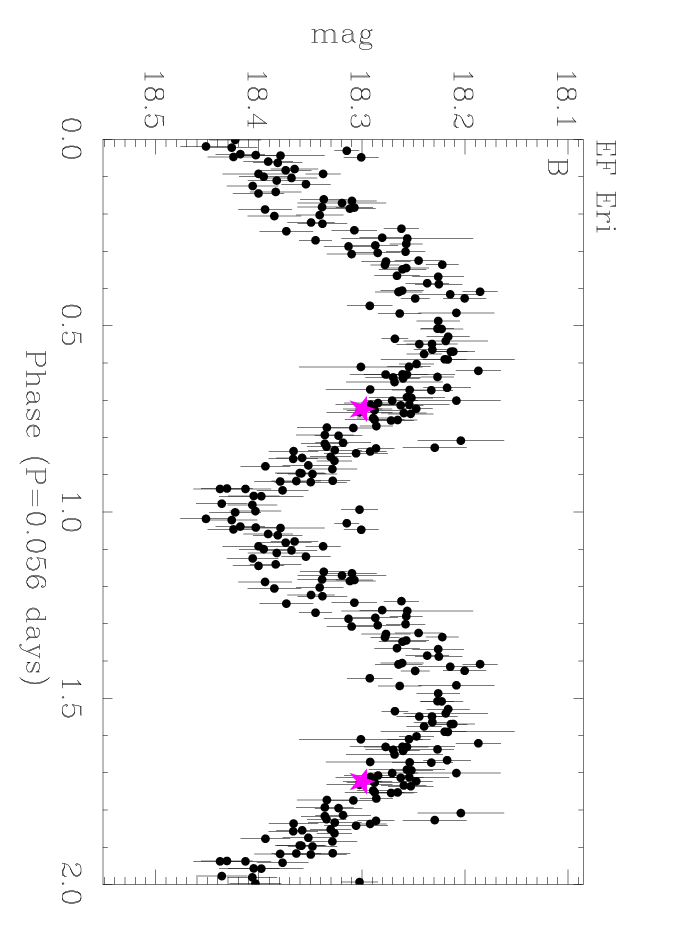}
\caption{SMARTS composite $B$ data for the extended low state during which
 the {\it HST} observations were made. The light curve is comprised
of 152 points on 68 nights from 2007 June 20 to 2008 Mar 23. The star near
phase 0.7 is the night of
2008 Jan 16.}
\end{figure}

\clearpage
\begin{figure} [h]
\figurenum {2}
\epsscale{1.0}
\includegraphics[width=15cm]{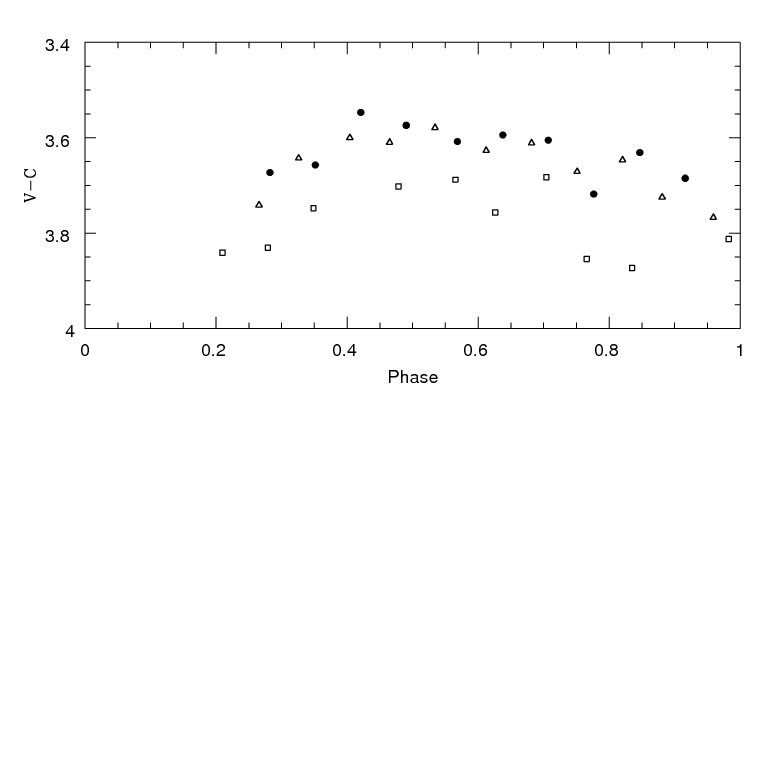}
\caption{NMSU $V$ data relative to a comparison star for Jan 15 (dots), Jan
16 (triangles) and Jan 18 (squares). Error bars are $\pm$0.03-0.05 mag for
Jan 15 and 16 and $\pm$0.06-0.08 mag on Jan 18.}
\end{figure}

\clearpage
\begin{figure} [h]
\figurenum {3}
\includegraphics[width=15cm]{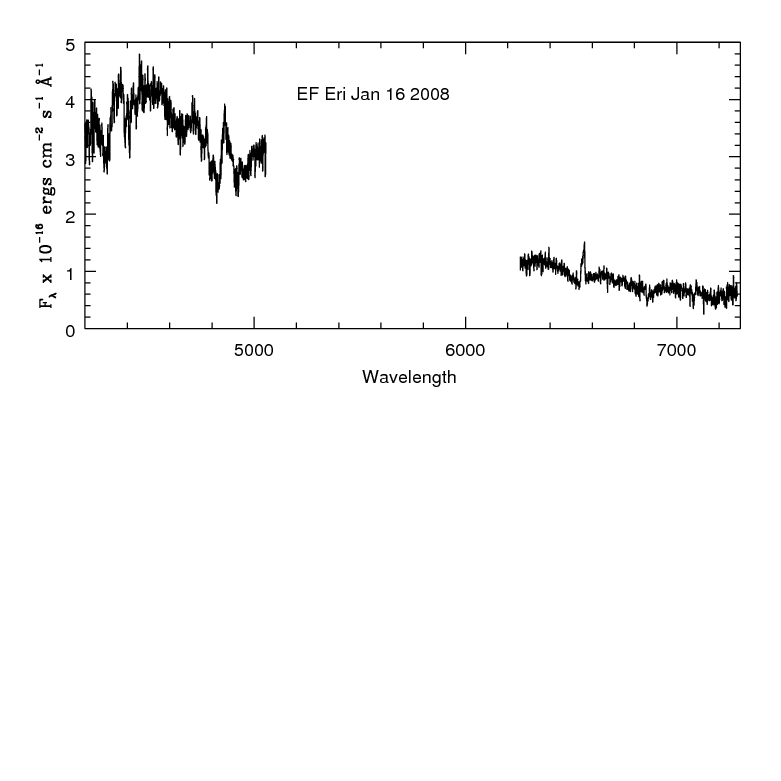}
\caption{DIS blue and red spectra averaged over the orbit of EF Eri. The Balmer
emission shows that some level of activity on the secondary star is present.}
\end{figure}

\clearpage
\begin{figure} [h]
\figurenum {4}
\includegraphics[width=15cm]{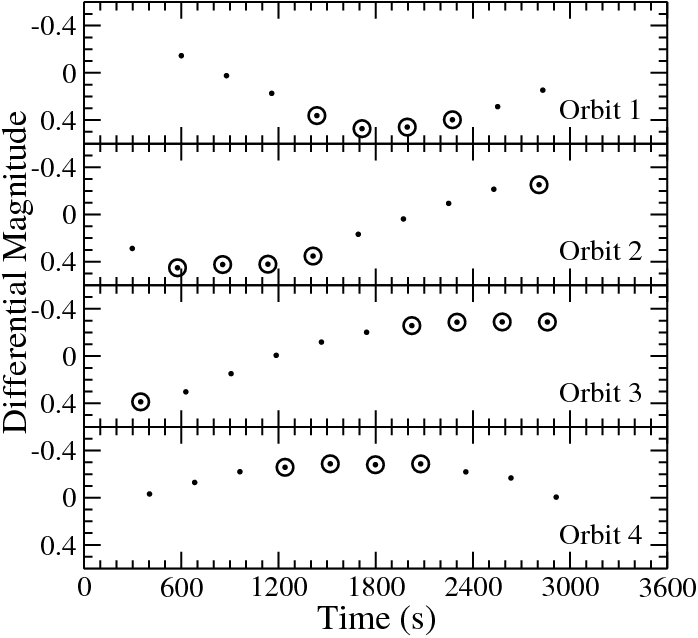}
\caption{Variability of EF Eri throughout each {\it HST} orbit. Each point is
a 239s integration, with a differential magnitude computed by dividing
the integrated flux by the average flux over all 4 orbits and converting
to a magnitude scale. The 9 circled points near minimum light are those used for 
the minimum flux values (within 23\%), and the 9 circled points near 
maximum light are those within 23\% of the maximum flux values.}
\end{figure}

\clearpage
\begin{figure} [h]
\figurenum {5}
\includegraphics[width=15cm]{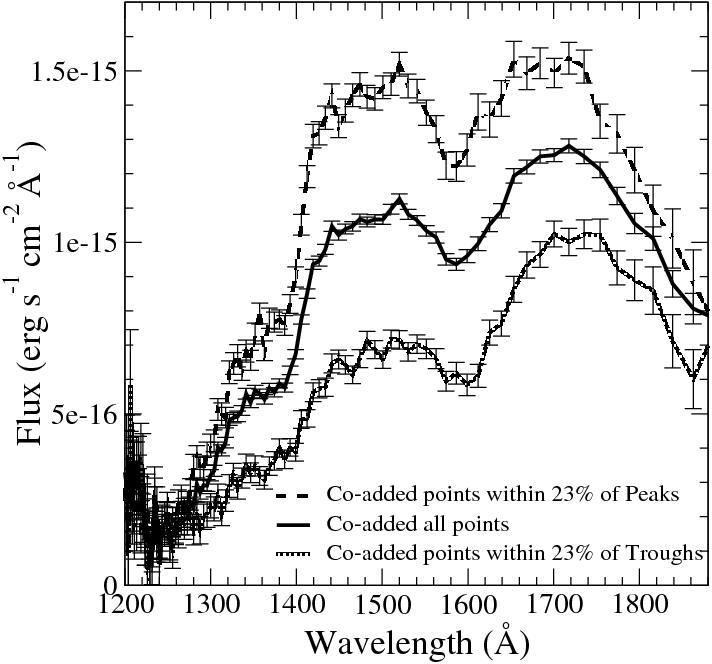}
\caption{Average spectrum with an extraction width of 17 using all 39 data points 
(middle solid line), with the 9 points within 23\% of the lowest flux points 
(bottom dotted line) and
with the 9 points within 23\% of the highest flux points (top dashed line).}
\end{figure}

\clearpage
\begin{figure} [h]
\figurenum {6}
\includegraphics[angle=270,width=15cm]{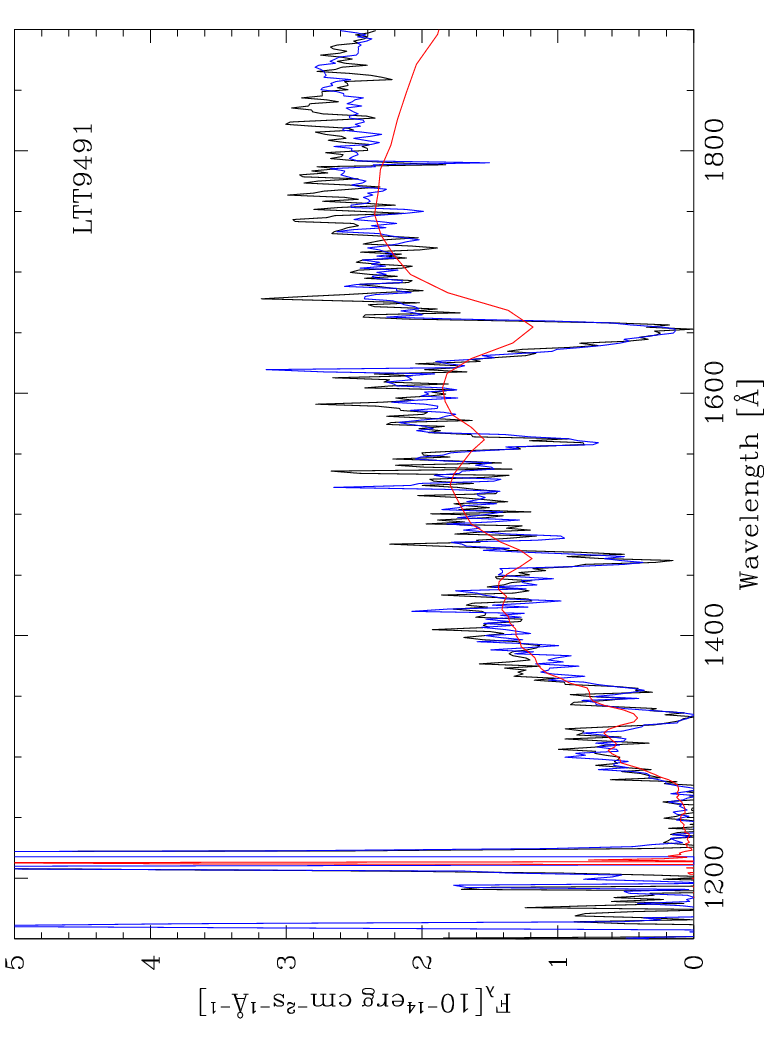}
\caption{Two IUE spectra of LTT9491 (high resolution lines) compared to the SBC spectrum
(low resolution line). Continuum fluxes match well until 1750\AA\ where the SBC
flux declines relative to the IUE values by about 30\%.}
\end{figure}

\clearpage
\begin{figure} [h]
\figurenum {7}
\includegraphics[width=15cm]{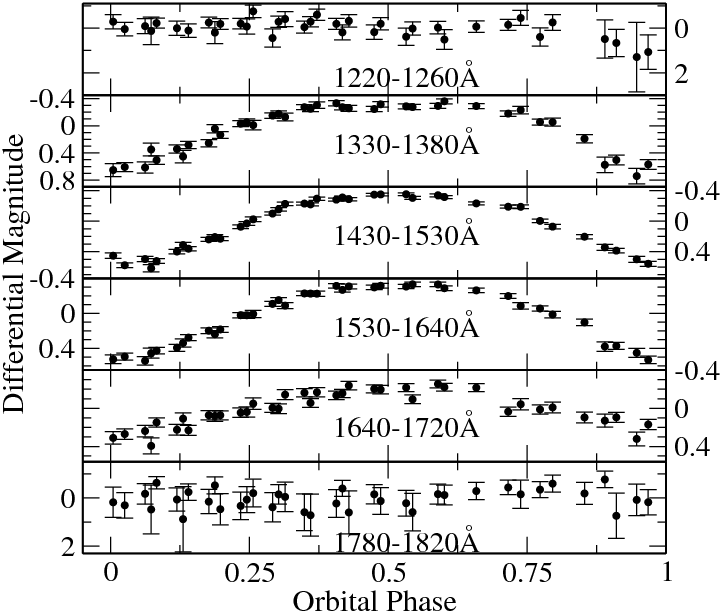}
\caption{Light curves in 6 UV bands as a function of spectroscopic orbital phase.
Each point is the integration of the flux within the bandpass.} 

\end{figure}

\clearpage
\begin{figure} [h]
\figurenum {8}
\includegraphics[angle=270,width=15cm]{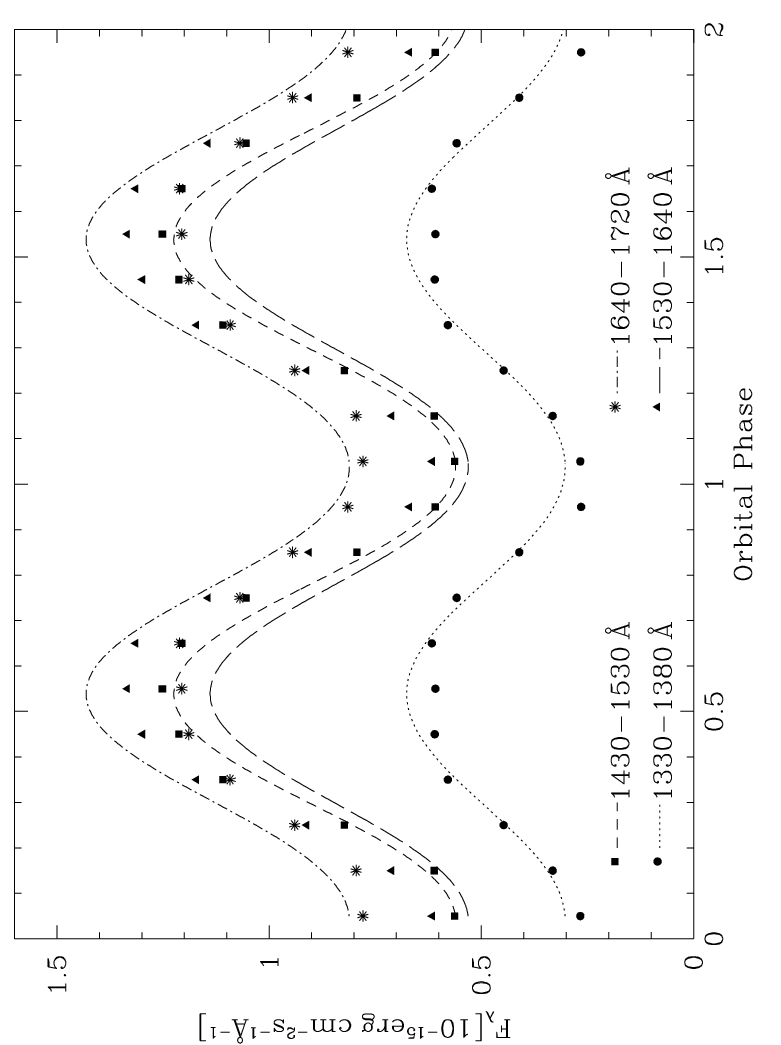}
\caption{
Best fit to the ACS/SBC light curves in four wavelength bands using
the white dwarf plus pole cap model. The observed ACS/SBC fluxes are
indicated by different symbols, the models corresponding to these
bands by different line styles.}
\end{figure}

\clearpage
\begin{figure} [h]
\figurenum {9}
\includegraphics[angle=270,width=15cm]{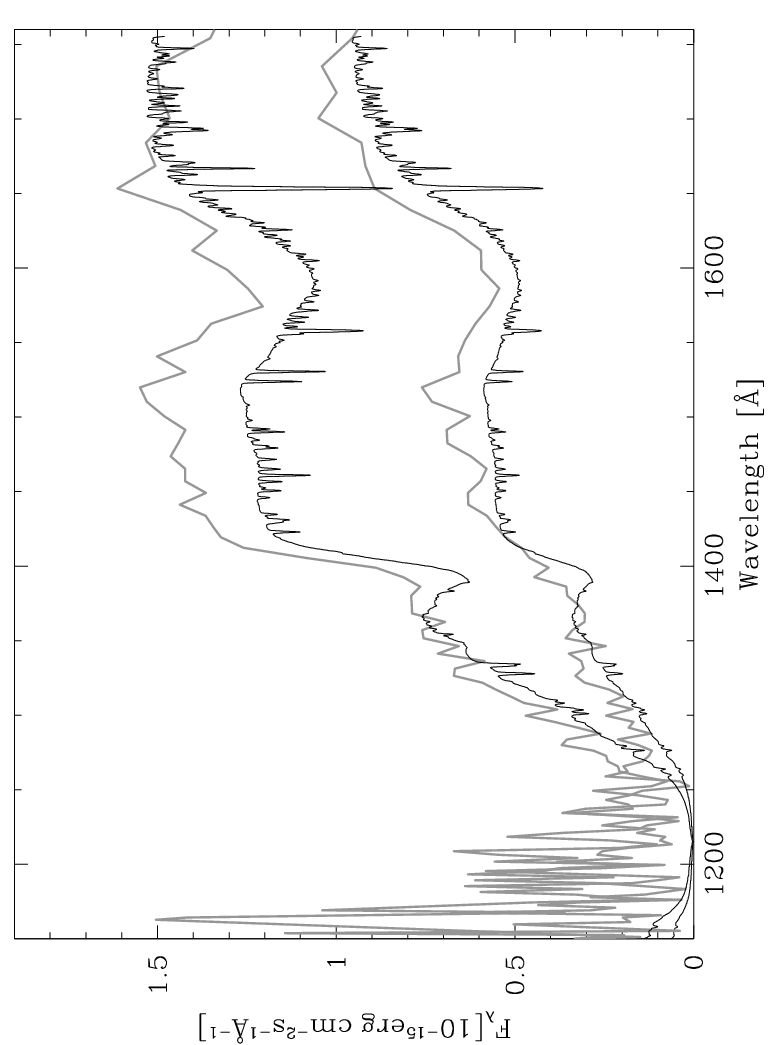}
\caption{
Observed (gray lines) and model spectra at orbital minimum (bottom
curves) and orbital maximum (top curves). The model spectra are
generated from the best-fit to the ACS/SBC light curves
(Figure 7). The broad depressions near
1400\,\AA\ and 1600\,\AA\ are the quasi-molecular H$_{2}^{+}$ and H$_{2}$
absorptions typical of cool, high-gravity hydrogen atmospheres. }
\end{figure}

\clearpage
\begin{figure} [h]
\figurenum {10}
\includegraphics[width=12cm]{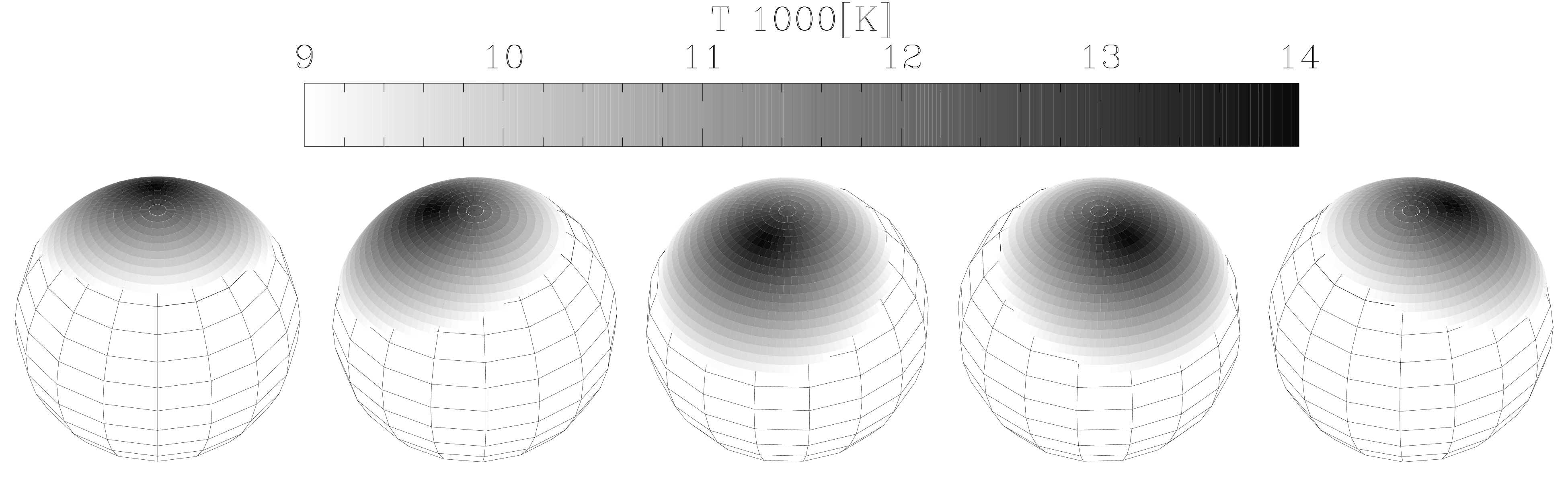}
\caption{Geometry of the heated pole cap in EF\,Eri. The unheated white dwarf
is represented by grid lines, the pole cap is gray-shaded. Orbital
phases are 0.1, 0.3, 0.5, 0.7, and 0.9 (from left to right).}
\end{figure}

\clearpage
\begin{figure} [h]
\figurenum {11}
\includegraphics[width=15cm]{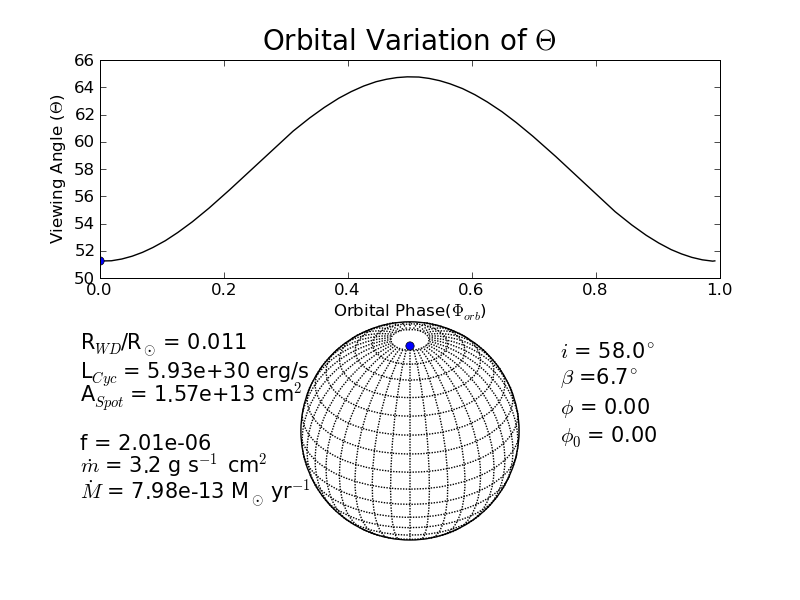}
\caption{Derived orbital geometry, accretion spot size and accretion
rates and the orbital variation of the viewing angle of the magnetic pole. $\phi$ is the current
orbital phase and $\phi_{0}$ is maximum light.}
\end{figure}

\clearpage
\begin{figure} [h]
\figurenum {12}
\includegraphics[width=15cm]{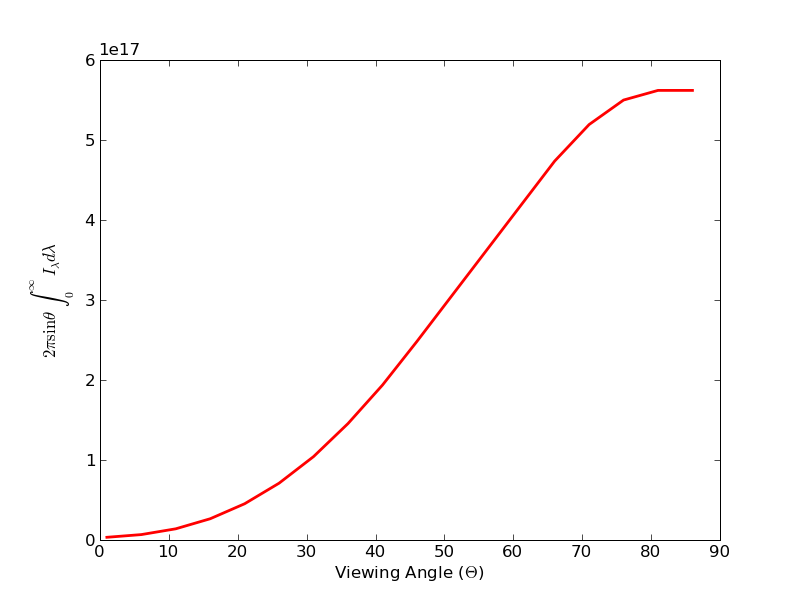}
\caption{Dependence of the total integrated intensity as a
function of the viewing angle $\Theta$. The area under the curve represents
the total cyclotron flux.}
\end{figure}

\clearpage
\begin{figure} [h]
\figurenum {13}
\includegraphics[height=4in,width=5.5in]{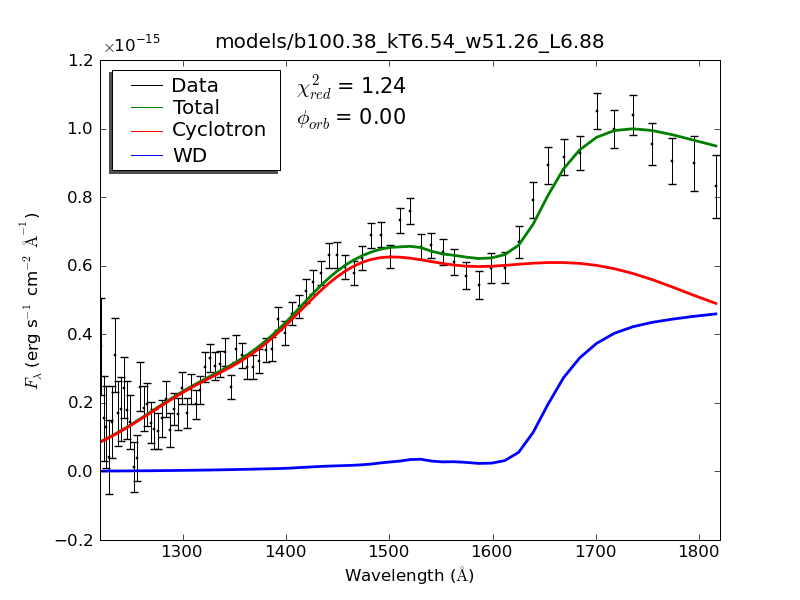}\newline
\includegraphics[height=4in,width=5.5in]{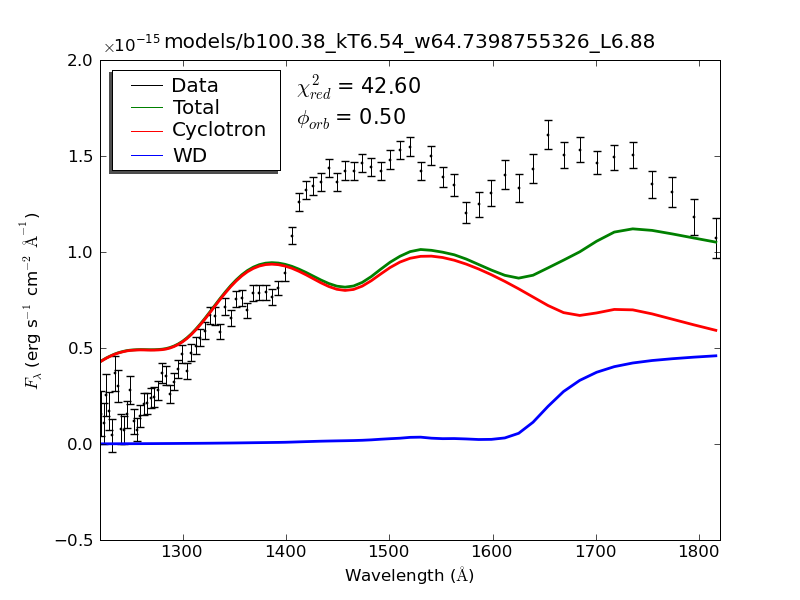}
\caption{Best fit model spectrum to the phase 0.0 {\it HST} spectrum (top)
and phase 0.5 {\it HST} spectrum (bottom).  Data are points
with error bars. Lower curves are a 9500K white dwarf, mid curves are the
cyclotron component for B=100 MG and top curves are the sum of the two
components.}
\end{figure}

\clearpage
\begin{figure} [t]
\figurenum {14}
\includegraphics[width=15cm]{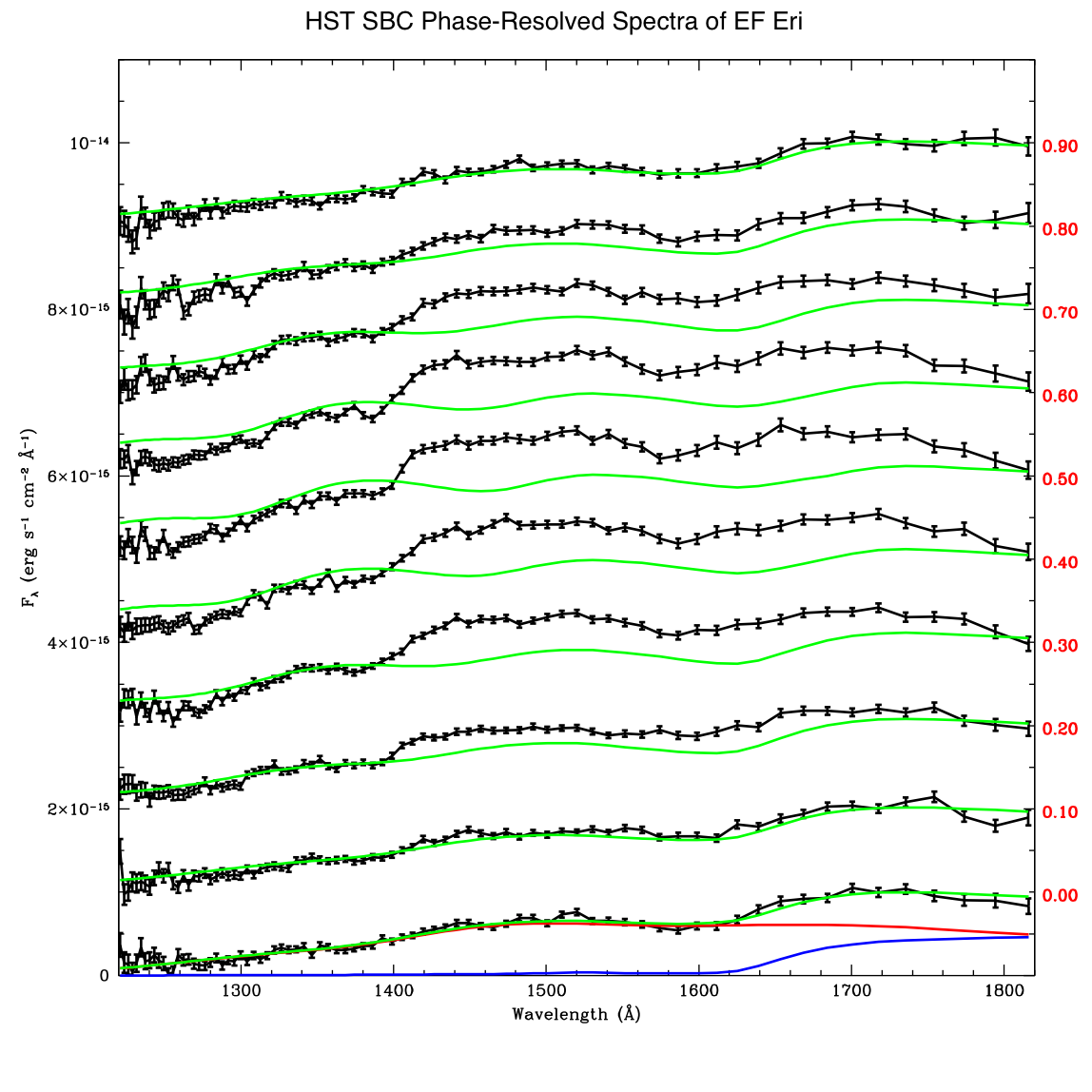}
\caption{Phase-resolved {\it HST} spectra (black with error bars) along with
the cyclotron (red), white dwarf (blue) and co-added CYC+WD model (green)
for the phase = 0.0 data. Each subsequent phase shows only the data
and co-added models. Fluxes at phase 0.0 are real, each subsequent
phase has a constant offset of 1$\times$10$^{-15}$ added to the
flux for clarity of presentation.}
\end{figure}

\clearpage
\begin{figure} [h]
\figurenum {15}
\includegraphics[width=15cm]{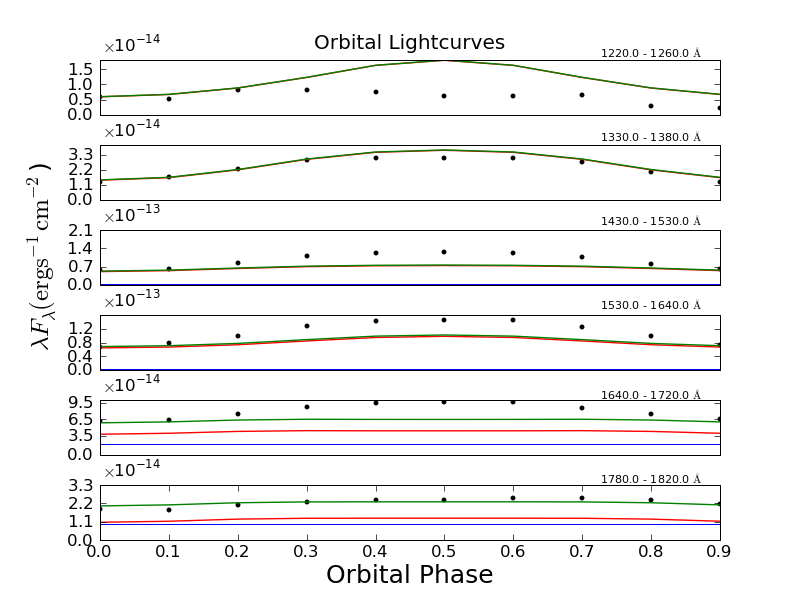}
\caption{Light curves in 6 bandpasses as a function of orbital phase,
with the white dwarf, cyclotron and sum model shown along with the
10 phase points created in each bandpass by integrating the fluxes within
each bandpass region listed on the right.}
\end{figure}

\clearpage
\begin{figure} [h]
\figurenum {16}
\includegraphics{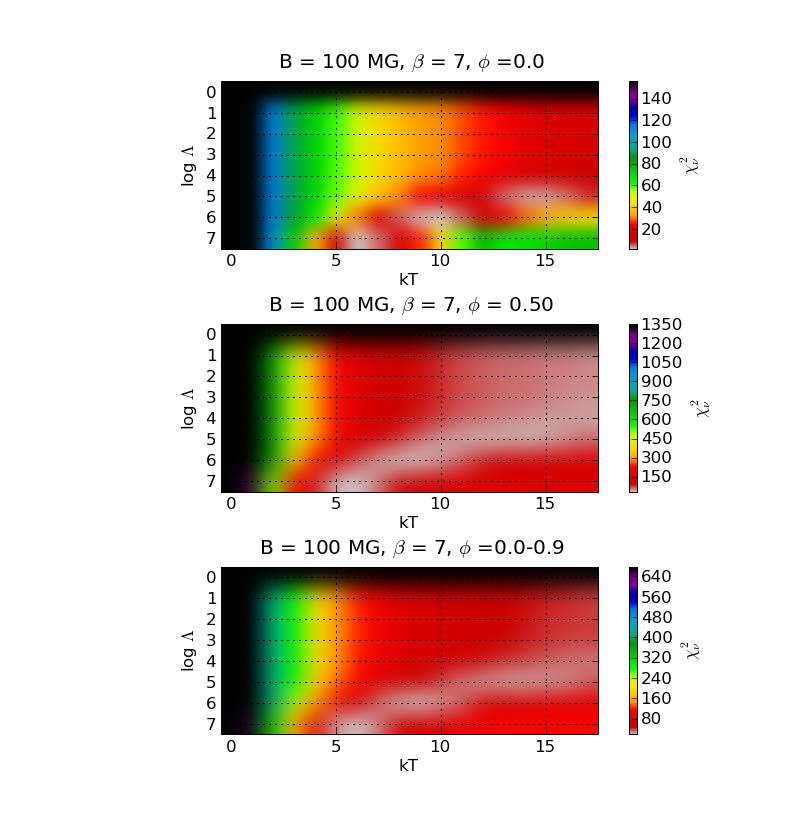}
\caption{$\chi^{2}$ heat maps of the kT-log$\Delta$ parameter space for
B=100 MG and $\beta$=7$^{\circ}$ at phases 0.0, 0.5 and the average over
the entire orbit.}
\end{figure}
 
\end{document}